\documentclass[11pt]{article}
\usepackage[margin=1.5in]{geometry}

\usepackage{algpseudocode}
\usepackage{algorithm}

\usepackage{graphicx}

\usepackage{array}
\usepackage{amsmath}
\usepackage{amssymb}

\usepackage{mathtools}

\usepackage{listings}
\usepackage{color}

\begin{document}

\title{An exact, cache-localized algorithm for the sub-quadratic
  convolution of hypercubes}

\author{Oliver Serang \\
	Freie Universit\"at Berlin and\\
        Leibniz-Institute of Freshwater Ecology and Inland Fisheries (IGB)\\
        orserang@uw.edu
}

\date{{\small 17 July, 2015}}

\maketitle

\begin{abstract}
\noindent Fast multidimensional convolution can be performed naively
in quadratic time and can often be performed more efficiently via the
Fourier transform; however, when the dimensionality is large, these
algorithms become more challenging. A method is proposed for
performing exact hypercube convolution in sub-quadratic time. The
method outperforms FFTPACK, called via {\tt numpy}, and FFTW, called
via {\tt pyfftw}) for hypercube convolution. Embeddings in hypercubes
can be paired with sub-quadratic hypercube convolution method to
construct sub-quadratic algorithms for variants of vector convolution.
\end{abstract}

\section*{Introduction}
\label{intro}
Convolution between tensors $x$ and $y$ is defined as 
\[
\left( x ~*~ y \right)\left[ k \right] = \sum_{ i,j ~:~ k=i+j } ~x[i] \cdot y[j],
\]
where $i$, $j$, and $k$ are integer vectors of the same
dimension. Multidimensional convolution is important for computing the
probability mass functions of sums and differences of joint
probability distributions (or equivalently, sums and differences of
multidimensional random variables) \cite{serang:probabilistic}.

\begin{algorithm}
  \caption{ {\bf Naive multidimensional convolution}}

  \label{algorithm:naive-convolution}
  \begin{small}
    \begin{algorithmic}[1]
      \Procedure{convolve}{$x$, $y$}
      
      \State $z \gets tensor(x.shape + y.shape - 1)$ \Comment{Initialize with zeros}

      \For{$i$ from $(0,0,\ldots)$ to $x.shape$}
      \For{$j$ from $(0,0,\ldots)$ to $y.shape$}
      \State $z[i+j] ~\mbox{+=}~ x[i] \cdot y[j]$
      \EndFor
      \EndFor

      \State {\bf return} $z$
      \EndProcedure
    \end{algorithmic}
  \end{small}
\end{algorithm}

The naive algorithm for multidimensional convolution simply performs
the Cartesian product between all $x[i]$ and all $y[j]$ ({\bf
  Algorithm}~\ref{algorithm:naive-convolution}). In practice, the
naive method is rarely used and is frequently eschewed in favor of
multidimensional fast Fourier transform (FFT) convolution. FFT
convolution is a numerical method, which typically uses Cooley-Tukey
method by padding each axis of $x$ and $y$ with zeros until it is a
power of $2$ and then further padding it with zeros until it reaches
the next power of $2$ \cite{cooley:algorithm}. Then the
multidimensional FFT of the padded $x$ and $y$ are performed (the
multidimensional FFT is equivalent to performing the 1D FFTs of every
row, column, \emph{etc.} of the input tensor), multiplied
element-wise, and then the multidimensional inverse FFT is performed
(this is the conjugate of the multidimensional FFT of the conjugate of
the input tensor). The convolution result will be found inside the
result of the multidimensional inverse FFT
\cite{proakis:introduction}.

For 1D vectors of length $N \gg 1$, the Cooley-Tukey FFT convolution
is substantially faster, requiring $O(N \log_2(N))$ steps, compared to
the $O(N^2)$ steps for the naive convolution based on the Cartesian
product. For 2D matrices of shape $(N_1, N_2)$, the row FFTs will cost
$O(N_1 \cdot N_2 \log_2(N_2))$ and the column FFTs will cost $O(N_2
\cdot N_1 \log_2(N_1))$ for an overall runtime in $O(N_1 \cdot N_2
(\log_2(N_1) + \log_2(N_2))) = O(N_1 \cdot N_2 \log_2(N_1 \cdot
N_2))$; therefore, FFT convolution still requires $O(N \log_2(N))$
steps where $N$ is the flat length of the 2D matrix; when $N$ is
large, this can be much faster than the $O(N^4)$ steps required by
naive convolution. However, for each problem listed here the
dimensionality is a constant, and is therefore not included in the
big-oh runtime. Furthermore, the column FFTs will not access
contiguous blocks of memory, and therefore will not cache efficiently
without special consideration.

But in the extreme case of convolving two $D$-dimensional hypercubes
$\{0,1\}^D$ (\emph{i.e.}, a tensor with $D$ axes, which each have
shape $\{0,1\}$), constants introduced by the dimensionality cannot be
ignored: While the flat length $N = 2^D$, zero padding doubles each
axis from $2$ values to $4$ values, and thus the zero padded tensor
has flat length $N=4^D$ or equivalently, $N={\left(2^2\right)}^D =
{\left(2^D\right)}^2$. Thus the runtime of the Cooley-Tukey FFT
convolution will be in $O\left({\left(2^D\right)}^2
\log_2({\left(2^D\right)}^2) \right) = O\left({\left(2^D\right)}^2
\cdot D \right) = O\left( N^2 \log_2(N) \right)$. In comparison, the
naive approach will require $O\left({ \left( 2^D \right)}^2 \right) =
O\left( N^2 \right)$ steps, and the runtime of convolving hypercubes
via the Cooley-Tukey FFT can be resemble or exceed the runtime of a
naive approach. Note that even though the multidimensional indices
will be vectors of $D$ booleans and require $D$ machine steps for each
step in the Cartesian product, these boolean indices can possibly be
embedded into a single machine precision integer.

In terms of the flat length, both naive and Cooley-Tukey convolution
of hypercubes will be in $\Omega(N^2)$. As an alternative to the
multidimensional Cooley-Tukey FFT, the hypercubes can be zero padded
into a $3^D$ tensor of the result size (rather than the $4^D$ tensor
used by the Cooley-Tukey method), and then the multidimensional FFT
can be performed as several row, column, \emph{etc.} 1D FFTs of length
$3$. These 1D FFTs cannot be performed via the Cooley-Tukey algorithm
because their lengths are not powers of $2$, but they can be performed
as DFTs in $3\times 3$ steps. That algorithm will require $O(3^D
\log_2(3^D))$ steps.

This manuscript presents an alternative algorithm for the convolution
of hypercubes. Unlike FFT convolution, this algorithm is exact and
visits the indices row order, and is thus inherently more cache
performant. The proposed method does not use FFT, but still achieves
the same sub-quadratic asymptotic runtime as the non-Cooley-Tukey FFT
of $3^D$ hypercubes.

The proposed method resembles a hypercube variant of Karatsuba's
method \cite{karatsuba:multiplication}; but where Karatsuba's method
is for performing multiplication of integers of arbitrary length, this
method is for convolving hypercubes of arbitrary dimension. Unlike
Karatsuba's method, these hypercubes can also contain real or floating
point values rather than integral values, because the recursion is
applied to hypercubes of smaller dimension rather than Karatsuba's
method of applying the recursion by splitting integral strings in
half. Also in contrast to Karatsuba's fast integer multiplication
method, the result space of two $N$-digit integers will have $2 \cdot
N-1$ digits, whereas the result space of hypercube convolution in
dimension $D$ will be $3^D$, which is in $\Omega(N^{1.585})$ (note
that the exponent has been rounded). This means that an algorithm with
the runtime of Karatsuba's method would be optimal.

The method can be used to simply and efficiently compute sums and
differences of multiple high-dimensional probability mass functions
where all axes are Bernoulli variables. Hypercube embeddings can be
used to efficiently solve alternate forms of 1D convolution.

\section*{Methods}
Observe that a tensor convolution of dimension $D$ can be performed by
peeling off the first axis to produce convolutions of dimension $D-1$:
\begin{eqnarray*}
z &=& x ~*~ y \\
\forall k_1, k_2, \ldots,~ z[k_1, k_2, \ldots] &=& \sum_{i_1, i_2, \ldots} x[i_1, i_2, \ldots] \cdot y[k_1-i_1, k_2-i_2, \ldots] \\
&=& \sum_{i_1} \sum_{i_2, \ldots} x[i_1, i_2, \ldots] \cdot y[k_1-i_1, k_2-i_2, \ldots] \\
&=& \sum_{i_1} x[i_1] ~*~ y[k_1-i_1],
\end{eqnarray*}
where tensors are stored in row-major format, so that $x[i_1]$ returns
a tensor of dimension $D-1$. From this principle, hypercube
convolution in dimension $D$ can be solved by four hypercube
convolutions of dimension $D-1$ as follows. 

As shown by the general tensor convolution case above, $k=i+j$
requires that the first index $k_1 = i_1 + j_1$. There are three cases
for $k_1$: $k_1 \in \{0,1,2\}$. $k_1 = 0$ requires $i_1=0,~
j_1=0$. Likewise, $k_1 = 2$ requires $i_1=1,~ j_1=1$. Lastly, $k_1 =
1$ can occur when either $i_1=0,~ j_1=1$ or $i_1=1,~ j_1=0$. Thus, $z
= x * y$ can be solved by performing 4 hypercube convolutions of
dimension $D-1$:
\begin{eqnarray*}
z[0] &=& x[0] * y[0] \\
z[1] &=& (x[0] * y[1]) + (x[1] * y[0]) \\
z[2] &=& x[1] * y[1],
\end{eqnarray*}
where each $*$ operation performs a hypercube convolution of dimension
$D-1$ and the base case simply returns the product between the two
numeric values when $D=0$. This leads to a runtime recurrence $T(D) =
4 \cdot T(D-1) + \Omega(2^D)$, which has closed form $T(D)$ in $\Omega\left(
{\left(2^D\right)}^2 \right)$, leaving an algorithm that is still
quadratic in $N$.

However, the fact that the tensors are hypercubes can be exploited:
First compute $z[0]$ and $z[2]$ as above. Then compute the
convolutions of marginals over the rows, $t = (x[0] + x[1]) * (y[0] +
y[1]) = x[0] * y[0] + x[0] * y[1] + x[1] * y[0] + x[1] * y[1]$. $t -
z[0] - z[2] = x[0] * y[1] + x[1] * y[0] = z[1]$. Thus, the $D$
dimensional convolution can be computed in only 3 convolutions of
dimension $D-1$. This will be exact when the dynamic range of the
values allows $(a + b) - a = b$.

The runtime of computing $x[0] + x[1]$ is $2^{D-1}$ (the same is true
for $y[0] + y[1]$) and the runtime of subtracting $t - z[0]$ is
$3^{D-1}$ (the same is true when subsequently subtracting $z[2]$);
therefore, the runtime recurrence is defined:
\begin{eqnarray*}
T(D) &=& 3 \cdot T(D-1) + 2 \cdot 2^{D-1} + 2 \cdot 3^{D-1}\\
&<& 3 \cdot T(D-1) + 2 \cdot 3^D\\
&=& 3 \cdot \left( 3 \cdot T(D-2) + 2 \cdot 3^{D-1} \right) + 2 \cdot 3^D\\
&\vdots&\\
&=& 3^D \cdot T(0) + 2 \cdot 3^D \cdot D \\
&=& 3^D + 2 \cdot 3^D \cdot D \\
&<& 3^{D+1} \cdot D.
\end{eqnarray*}
$N = 2^D$, so the runtime will be bounded by $3 \cdot D \cdot
3^{\log_2(N)} = 3 \cdot D \cdot N^{\log_2(3)} \leq 3 \cdot D \cdot
N^{1.585} \in O(N^{1.585} \log_2(N))$. All operations are performed by
pairing two contiguous blocks of memory (much like a 1D in-place
Cooley-Tukey FFT after bit-reversal has been performed), and so the
proposed method is highly cache performant ({\bf Algorithm 2}).

\begin{small}
  \lstset{language=C++,
    basicstyle=\ttfamily,
    keywordstyle=\color{blue}\ttfamily,
    stringstyle=\color{red}\ttfamily,
    commentstyle=\color{green}\ttfamily,
    morecomment=[l][\color{magenta}]{\#},
    breaklines=true
  }
  
\noindent\rule{\textwidth}{1.5pt}
\begin{lstlisting}[title={{\bf Algorithm 2 Fast hypercube convolution in C++.} The algorithm is invoked {\tt HypercubeConvolution<D>::apply(dest, x, y, flat\_length\_of\_dest)}. The contents of {\tt x} and {\tt y} will be modified as the algorithm runs.}]
template <unsigned int D>
class HypercubeConvolution {
public:
  static void apply(double*__restrict__ const dest, double*__restrict__ const x, double*__restrict__ const y, const unsigned long three_to_d) {
    const unsigned long two_to_d_minus_one = 1ul<<(D-1);

    // Compute dest[0]:
    HypercubeConvolution<D-1>::apply(dest, x, y, three_to_d/3);

    // Compute dest[2]:
    HypercubeConvolution<D-1>::apply(dest + 2*three_to_d/3, x + two_to_d_minus_one, y + two_to_d_minus_one, three_to_d/3);

    // Compute x[0] + x[1] and y[0] + y[1]:
    unsigned int k;
    for (k=0; k<two_to_d_minus_one; ++k)
      x[k] += x[k + two_to_d_minus_one];

    for (k=0; k<two_to_d_minus_one; ++k)
      y[k] += y[k + two_to_d_minus_one];

    // Compute dest[1] = conv(x[0] + x[1], y[0] + y[1]) - dest[0] - dest[2]
    HypercubeConvolution<D-1>::apply(dest + three_to_d/3, x, y, three_to_d/3);
    for (k=0; k<three_to_d/3; ++k)
      dest[k + three_to_d/3] -= dest[k];
    for (k=0; k<three_to_d/3; ++k)
      dest[k + three_to_d/3] -= dest[k + 2*three_to_d/3];
  }
};

template <>
class HypercubeConvolution<1u> {
public:
  static void apply(double*__restrict__ const dest, double*__restrict__ const x, double*__restrict__ const y, unsigned long) {
    dest[0] = x[0] * y[0];
    dest[1] = x[1] * y[0] + x[0] * y[1];
    dest[2] = x[1] * y[1];
  }
};
  \end{lstlisting}
\end{small}
\noindent\rule{\textwidth}{0.5pt}

\section*{Results}
The runtime and accuracy of the divide and conquer method is compared
to Python's Cooley-Tukey-based {\tt fftconvolve} routine from {\tt
  scipy.signal} and to convolution via Python's {\tt numpy.fft.fftn}
to tensors of the result shape $3^D$, which are both implemented in
Fortran by {\tt numpy}. The runtime and accuracy are also compared to
FFTW on $3^D$ tensors via the {\tt pyfftw} package
\cite{frigo:fast}. The proposed divide and conquer algorithm is
implemented without complex numbers or any libraries using $<40$ lines
of template recursive C++ code. The template recursive formulation is
easily optimized, because the recursive calls can be unrolled and
inlined by the compiler (it is compiled with {\tt clang++-3.8 \tt
  -Ofast}). For each $D$, three simulations were performed and the
median runtime is reported ({\bf Table}~\ref{table:runtimes}). FFTW is
first run without timing it so that the expensive {\tt plan} step
(essentially a form of just-in-time compilation) used by FFTW is not
included in the runtime.

\begin{table}[ht!]
\centering
\small
\scalebox{0.78}{
\begin{tabular}{r|rrrrrrrr}
\hline
$D$ & $11$ & $12$ & $13$ & $14$ & $15$ & $16$ & $17$ & $18$ \\
$N$ & $2048$ & $4096$ & $8192$ & $16384$ & $32768$ & $65536$ & $131072$ & $262144$ \\
\hline \\
& \multicolumn{8}{c}{Runtimes (seconds)} \\
\hline \\
{\tt fftconvolve} & 0.930 & 4.00 & 17.1 & 81.0 & ----- & ----- & ----- & -----  \\
{\tt fftn} of $3^D$ & 0.078 & 0.266 & 0.902 & 3.19 & 10.2 & 35.5 & 115 & ----- \\
{\tt FFTW} of $3^D$ & 0.0136 & 0.0358 & 0.181 & 0.420 & 1.89 & 4.19 & 13.4 & ----- \\
D\&C & {\bf 0.00447} & {\bf 0.0155} & {\bf 0.0396} & {\bf 0.0429} & {\bf 0.167} & {\bf 0.444} & {\bf 1.39} & {\bf 4.56} \\
\hline \\
& \multicolumn{8}{c}{Relative error at smallest element} \\
\hline \\
{\tt fftconvolve} & {\bf 0} & {\bf 0} & {\bf 0} & {\bf 0} & ----- & ----- & ----- & -----  \\
{\tt fftn} of $3^D$ & $1\cdot 10^{-8}$ & $1.7\cdot 10^{-7}$ & $1.08\cdot 10^{-6}$ & $1.52\cdot 10^{-5}$ & $1.23\cdot 10^{-4}$ & $3.95\cdot 10^{-4}$ & $2.98\cdot 10^{-3}$ & ----- \\
{\tt FFTW} of $3^D$ & $4\cdot 10^{-8}$ & $1.8\cdot 10^{-7}$ & $1.32\cdot 10^{-6}$ & $1.52\cdot 10^{-5}$ & $1.76\cdot 10^{-5}$ & $6.62\cdot 10^{-4}$ & $8.35\cdot 10^{-3}$ & ----- \\
D\&C & {\bf 0} & {\bf 0} & {\bf 0} & {\bf 0} & {\bf 0} & {\bf 0} & {\bf 0} & {\bf 0} \\
\end{tabular}
}
\caption{{\bf Runtimes and accuracies on hypercube convolutions of
    dimension $D$.}  Runtimes (in seconds) computed using the {\tt
    time.time} routine in Python and {\tt std::clock\_t} time in
  C++. Note that {\tt fftconvolve} and {\tt fftn} both call {\tt
    FFTPACK} in Fortran, and that {\tt fftconvolve} is able to easily
  exploit the fact that the convolution is on the reals because it
  operates on the $4 \times 4 \times \cdots$ tensor (whose axes are
  powers of $2$). The {\tt fftn} $3^D$ method only zero pads into a
  $3\times 3\times \cdots$ tensor. FFTW uses multiple cores. All FFT
  methods perform row, column, \emph{etc.} DFTs to perform the forward
  and inverse tensor FFTs and numerically compute the
  convolution. Values are unreported when more than the $16$GB RAM
  available was required. Accuracy is evaluated using hypercubes with
  flat vectors $x.flat = y.flat = [1,2,3,\ldots 2^D]$ and analyzing
  the relative error at the smallest result value $(x ~*~
  y)[(0,0,\ldots 0)] = 1$, which can suffer the greatest influence
  from large values elements in the convolution. Note that because
  {\tt fftconvolve} is run on $4\times 4 \times \cdots$ tensors, the
  twiddle factors are computed using angles of the form $k \cdot
  \frac{\pi}{2}, ~k \in \mathbb{Z}$ and thus accumulate no error,
  whereas applying FFTs to $3\times 3\times \cdots$ tensors yields
  larger errors from angles of the form $k \cdot \frac{\pi}{3}$. Even
  though it uses only one core, the proposed divide and conquer
  algorithm (labeled D\&C) simultaneously achieves the highest
  accuracy and the fastest runtime (best results in bold font).}
\label{table:runtimes}
\end{table}

\section*{Discussion}
The resulting divide and conquer algorithm produces the same results
as with FFT convolution, but with exact results and with a
significantly faster runtime and lower memory footprint. Furthermore,
the increased precision of the proposed algorithm means it can be
paired with $p-$norm rings to more accurately approximate fast
algorithms on the semiring $(\times, \max)$ (\emph{e.g.},
max-convolution), because much greater values of $p$ will be
numerically stable \cite{serang:fast}. This means that max-convolution
on hypercubes containing integers of bounded dynamic range can be
performed exactly by using a value of $p$ large enough that the
relative error $1 - N^{\frac{-1}{p}}$ drops low enough that the
absolute error on the range of integers is $<0.5$, and so rounding
$p-$norm estimates to the nearest integer will be exact
\cite{pfeuffer:bounded}.

Hypercube embeddings can be used to perform variants of 1D vector
convolution. For example, the proposed hypercube convolution method
solves the 1D ``carry-free convolution''. Where 1D convolution defines
$z[k] = \sum_{i,j ~:~ k=i+j} ~x[i] \cdot y[j]$, the carry-free
convolution would describe the same problem where the additions
between integers apply no bitwise carry operations. For instance,
$i=7$ and $j=5$ would add bitwise $(1,1,1) + (1,0,1)$ to produce
$(2,1,2)$, which would produce $(1,1,0,0)$ after carry operations and
is equivalent to $8+4 = 12 = 7+5$. In the carry-free variant, the
carry operations would not be performed during convolution; therefore,
$x[(1,1,1)] \cdot y[(1,0,1)]$ would be added to $z[(2,1,2)]$ rather
than to $z[(1,1,0,0)]$. This seemingly innocuous change to 1D
convolution means that it is not trivial to solve it efficiently using
existing 1D convolution algorithms, but it can be solved by embedding
into a hypercube of dimension $D = \log_2(N)$, where each axis
corresponds to a $\{0, 1\}$ value of the bits for a given index, and
then convolving the two hypercubes. That hypercube convolution can be
efficiently and accurately solved by the method proposed. It is likely
that there are similar embeddings for other combinatoric problems that
would benefit.


\end{document}